\documentstyle[aps,floats,epsf]{revtex}

\begin{document}
\title
{Narrow $J^\pi=1/2^+$, $3/2^+$,  and $3/2^-$ states 
of $\Theta^+$ in a Quark model with Antisymmetrized Molecular Dynamics}

\author{Y. Kanada-En'yo, O. Morimatsu, and T. Nishikawa}

\address{Institute of Particle and Nuclear Studies, \\
High Energy Accelerator Research Organization,\\
1-1 Oho, Tsukuba, Ibaraki 305-0801, Japan}

\maketitle

\begin{abstract}
The exotic baryon $\Theta^+(uudd\bar{s})$ is studied with microscopic 
calculations in a quark model 
by using a method of antisymmetrized molecular dynamics.
We predict that three narrow states, $J^\pi=1/2^+(I=0)$, 
$J^\pi=3/2^+(I=0)$, and $J^\pi=3/2^-(I=1)$
nearly degenerate with the lowest $1/2^-$ state in the $uudd\bar{s}$ system.
We discuss $KN$ decay widths and estimate them to be $\Gamma< 7$ for 
the $J^\pi=\{1/2^+,3/2^+\}$, and $\Gamma<1$ MeV for the $J^\pi=3/2^-$ state.
In contrast to these narrow states, 
the $1/2^-$ states should be much broader.
We assign the observed $\Theta^+$ as the $J^\pi=\{1/2^+,3/2^+\}$.
\end{abstract}


\section{Introduction}

The exotic baryon $\Theta^+$ has recently been reported 
by several experimental 
groups\cite{leps,diana,clasa,clasb,saphir,itep,hermes,itep-2,zeus}. 
Since the quantum numbers determined from its decay modes 
indicate that the minimal quark content is $uudd\bar{s}$, 
these induced experimental and theoretical 
studies of multiquark hadrons. However it should be kept in mind that
the $\Theta^+$ has not been 
well established yet because of the low statistics and 
experimental reports\cite{bes,hera-b,phenix} for no evidence of 
the $\Theta^+$. 

The prediction of a $J^\pi=1/2^+$ state of $uudd\bar{s}$ 
by a chiral soliton model \cite{diakonov} motivated the experiments of the 
first observation of $\Theta^+$\cite{leps}. 
Their prediction of even parity is unnatural in the
naive quark model, because the lowest $q^4\bar{q}$ state 
is expected to be spatially symmetric and have odd parity due to the 
odd intrinsic parity of the anti-quark.
Theoretical studies were done to describe $\Theta^+$ 
by many groups\cite{jaffe,capstick,karliner,hosaka,oka,sasaki,fodor,jennings},
some of which predicted the opposite parity, 
$J^\pi=1/2^-$\cite{oka,sasaki,fodor}.
The problem of spin and parity of $\Theta^+$ is not only open but
also essential to understand the dynamics of pentaquark systems. 
To solve this problem, it is crucial to calculate five-quark system 
relying on less {\it a priori} assumptions
such as the existence of quark clusters or the spin parity. 

In this paper we would like to clarify the mechanism 
of the existence of the pentaquark baryon and predict possible narrow
$\Theta^+$ states. 
We try to extract a simple picture for the pentaquark 
baryon with its energy, width, spin, parity and also 
its shape from explicit 5-body calculation.
In order to achieve this goal, we study the pentaquark 
with a flux-tube model\cite{carlson,morimatsu} based on strong coupling QCD,
by using a method of antisymmetrized 
molecular dynamics (AMD)\cite{ENYObc,AMDrev}.
In the flux-tube model, the interaction energy of quarks 
and anti-quarks is given by the energy of the string-like 
color-electric flux, which is proportional to the minimal 
length of the flux-tube connecting quarks and anti-quarks 
at long distances supplemented by perturbative one-gluon-exchange (OGE) 
interaction at short distances.
For the $q^4\bar q$ system the flux-tube configuration 
has an exotic topology, Fig.\ref{fig:flux}(c), 
in addition to an ordinary meson-baryon
topology, Fig. \ref{fig:flux}(d), and the transition between different 
topologies takes place only in higher order of  the 
strong coupling expansion.
Therefore, it seems quite natural that the flux-tube model accommodates 
the pentaquark baryon.
In 1991, Carlson and Pandharipande studied exotic 
hadrons in the flux-tube model\cite{carlsonb}.
They calculated for only a few $q^4\bar q$ states with very limited quantum
numbers and concluded that pentaquark baryons are absent.
We apply the AMD method to the flux-tube model.
The AMD is a variational method to solve a finite many-fermion system.
This method is powerful for the study of nuclear structure.
One of the advantages of this method is that 
the spatial and spin degrees of freedom  
for all particles are independently treated.
This method can successfully describe various types of structure such as 
shell-model-like structure and clustering (correlated nucleons)
in nuclear physics. 
In the application of this method to a quark model,
we take the dominant terms 
of OGE potential and string potential 
due to the gluon flux tube.
Different flux-tube configurations are assumed to be decoupled.
Since we are interested in the narrow states, we only adopt the
confined configuration given by Fig.\ref{fig:flux}(c).
We calculate all the possible spin parity states of $uudd\bar{s}$
system, and predict low-lying states.
By analysing the wave function, we discuss the properties of $\Theta^+$
and estimate the decay widths of these states with a 
method of reduced width amplitudes.  

This paper is organized as follows.
We explain the formulation of the present framework in the next section, 
and show the results in \ref{sec:results}.
In \ref{sec:discuss}, 
we discuss the structure of low-lying states and their widths.
Finally, we give a summary in \ref{sec:summary}.

\section{formulation}\label{sec:framework}

In the present calculation, the quarks are treated as 
non-relativistic spin-$\frac{1}{2}$ Fermions.
We use a Hamiltonian as follows,
\begin{equation}
H=H_0+H_I+H_f,
\end{equation}
where $H_0$ is the kinetic energy of 
the quarks, $H_I$
represents the short-range OGE interaction between the quarks
and $H_f$ is the energy of the flux tubes.
For simplicity, we take into account 
the mass difference between the
$ud$ quarks and the $s$ quark, only 
in the mass term of $H_0$ but not in the kinetic energy term.
Then, $H_0$ is represented as follows;
\begin{equation}
H_0=\sum^{N_q}_{i} m_i+\sum^{N_q}_{i}\frac{{p}_i^2}{2m_q}-T_0,
\end{equation}
where $N_q$ is the total number of quarks and $m_i$ is the mass of 
$i$-th quark, which is $m_q$ for a $u$ or $d$ 
quark and $m_s$ for a $\bar{s}$ quark.
$T_0$ denotes the kinetic energy of the center-of-mass motion.

$H_I$ represents the short-range OGE interaction between quarks
and consists of the Coulomb and the 
color-magnetic terms,
\begin{equation}
H_I=\alpha_c\sum_{i<j}F_i F_j
\left[\frac{1}{r_{ij}}-\frac{2\pi}{3m_i m_j}s(r_{ij})
\sigma_i\cdot\sigma_j\right].
\end{equation}
Here, $\alpha_c$ is the quark-gluon coupling constant, and 
$F_iF_j$ is defined by 
$\sum_{\alpha=1,\cdots,8} F^\alpha_i F^\alpha_j$,
where $F^\alpha_i$ is the 
generator of color $SU(3)$, $\frac{1}{2}\lambda^\alpha_i$ for quarks and
$-\frac{1}{2}(\lambda^\alpha_i)^*$ for anti-quarks.
The usual $\delta(r_{ij})$ function in the spin-spin interaction is replaced by
a finite-range Gaussian, $s(r_{ij})
=\left[ \frac{1}{2\sqrt{\pi}\Lambda}\right]^3\exp{
\left[-\frac{r_{ij}^2}{4\Lambda^2}\right]}$, as in Ref.\cite{carlsonb}.
Of course, the full OGE interaction contains other terms 
such as tensor and spin-orbit interactions.
However, since our main interest here is to see the basic properties
of the pentaquark,
we do not include these minor contributions.

In the flux-tube quark model \cite{carlson},
the confining string potential is written as
$H_{f}={\sigma}L_f-M^0$,
where ${\sigma}$ is the string tension, $L_f$ is
the minimum length of the flux tubes, and $M^0$ is the 
zero-point string energy. 
$M^0$ depends on the topology of the flux tubes 
and is necessary to fit the $q\bar{q}$, $q^3$ and $q^4\bar{q}$ potential 
obtained from lattice QCD or phenomenology. In the present calculation, we 
adjust the $M^0$ to fit the absolute masses 
for each of three-quark and pentaquark.

\begin{figure}
\noindent
\epsfxsize=0.45\textwidth
\centerline{\epsffile{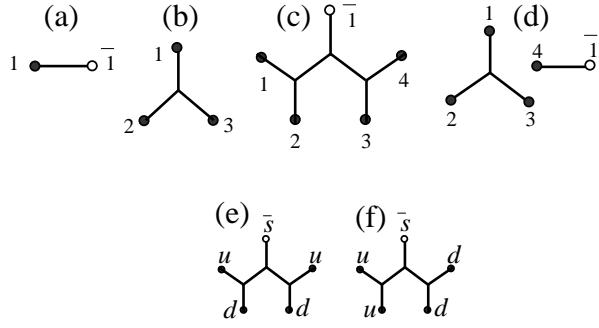}}
\ \\
\caption{\label{fig:flux}
Flux-tube configurations for confined states of $q\bar{q}$ (a),
$q^3$ (b), $q^4\bar{q}$ (c), and disconnected flux-tube of
$q^4\bar{q}$ (d). Figures (e) and (f) represent the flux tubes
in the color configurations, 
$[ud][ud]\bar{s}$ and
$[uu][dd]\bar{s}$, respectively.}
\end{figure}

\begin{figure}
\noindent
\epsfxsize=0.6\textwidth
\centerline{\epsffile{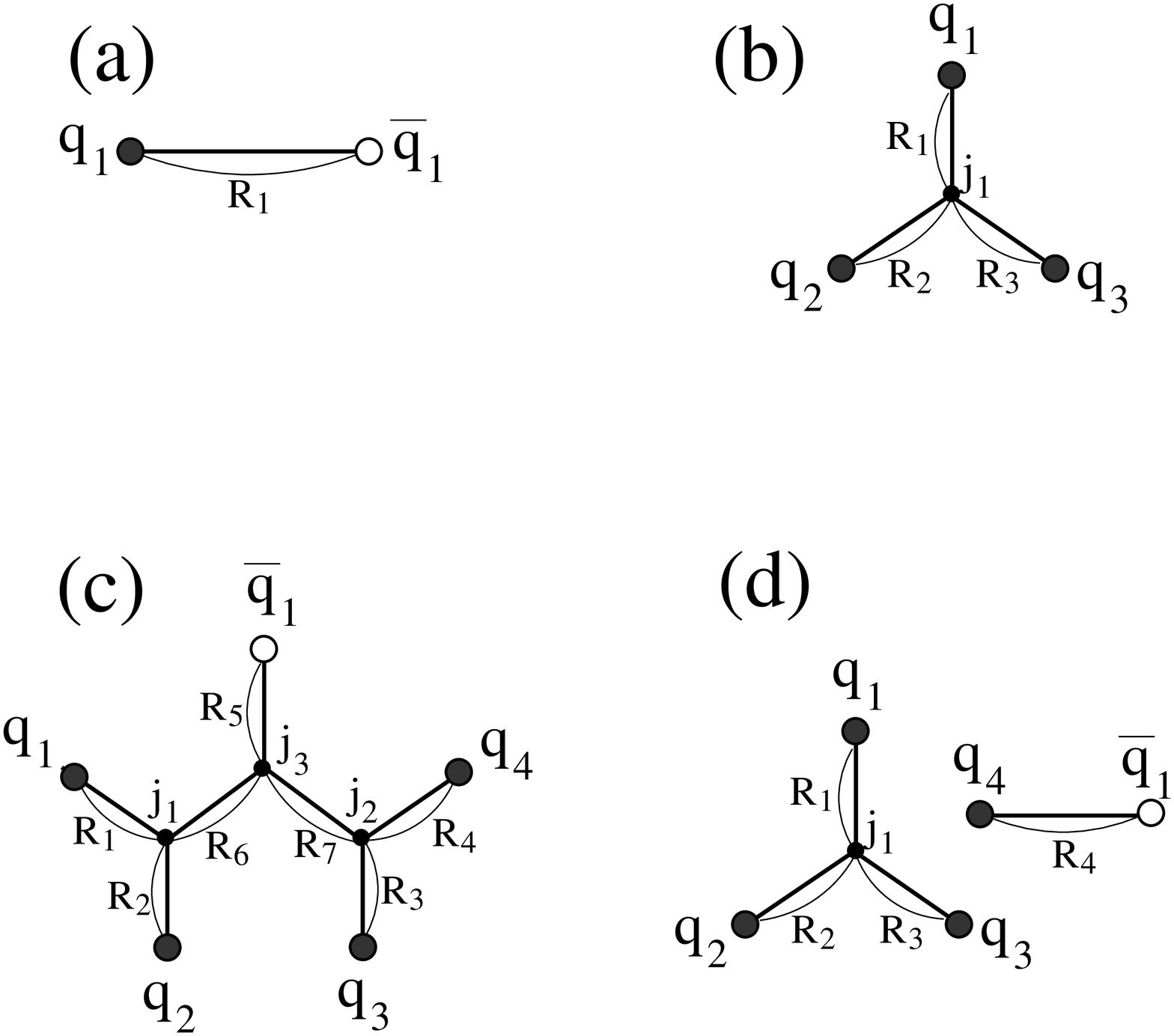}}
\ \\
\caption{\label{fig:flux-app}
Flux-tube topologies for the $q\bar{q}$ (a),
$q^3$ (b), $[q_1q_2][q_3q_4]\bar{q_1}$ (c), and disconnected flux tubes (d)
for the $(qqq)_1(q\bar{q})_1$.
The flux-tube topologies are described by the bonds with 
the lengths $R_k$ and the junctions $j_k$.}
\end{figure}

For the meson and 3$q$-baryon systems, the flux-tube configurations 
are the linear line and the $Y$-type configuration with three bonds and 
one junction as shown in Fig.\ref{fig:flux-app}(a) and (b), respectively. 
The string potential given by the $Y$-type flux tube in a $3q$-baryon
system is supported by Lattice QCD \cite{Takahashi}.
For the pentaquark system, the different types of 
flux-tube configurations 
appear as shown in Fig. \ref{fig:flux}.(e),(f), and (d), which correspond
to the states,
$|\Phi_{(e)}\rangle=|[ud][ud]\bar{s}\rangle$,
$|\Phi_{(f)}\rangle=|[uu][dd]\bar{s}\rangle$,
and $|\Phi_{(d)}\rangle=|(qqq)_{1}(qq)_1\rangle$,
respectively. ($[qq]$ is defined by color anti-triplet of $qq$.) 
The flux-tube configuration (e) or (f) have seven bonds and 
three junctions, while the configuration (d) has four bonds and one
junction.
In principle, besides these color configurations 
($[qq][qq]\bar{q}$ and $(qqq)_{1}(qq)_1$), other color configurations
are possible in totally color-singlet $q^4\bar{q}$ systems by incorporating
a color-symmetric $(qq)_6$ pair as in Refs.\cite{karliner,jennings}.  
However, since such a string from the $(qq)_6$ is energetically excited and is 
unfavored in the strong coupling limit of gauge theories as shown in 
Ref.\cite{Kogut75}. Therefore, 
we consider only color-$3$ flux tubes 
as the elementary tubes. In fact, the string tension for the color-6 string
in the strong coupling limit is 
$5/2$ times larger than that for the color-$3$ 
string from the expectation value of the Casimir operator.
The string potentials given by the tube lengths of the configuration
Fig.\ref{fig:flux}(c) is supported by Lattice QCD calculations
\cite{okiharu-5q}.

In the present calculation of the energy, we neglect the transition 
among $|\Phi_{(e)}\rangle$, $|\Phi_{(f)}\rangle$ and $|\Phi_{(d)}\rangle$
because they have different flux-tube configurations.
It is reasonable 
in the first order approximation, as mentioned before.
In each tube configuration, the minimum length $L_f$ is given by a
sum of the lengths($R_i$) of bonds $L_f=R_1+\cdots+R_k$
($k$ is the number of the bonds. See Fig.\ref{fig:flux-app}).
Here we define $L_{ij}$ to be the length  of the path between 
$i$-th (anti)quark 
and $j$-th (anti)quark along the flux tubes.
For example, in case of the $[qq][qq]\bar{q}$ state shown in 
Fig.\ref{fig:flux-app}(c), 
the path lengths are given by the bond lengths $R_i$ as 
$L_{12}=R_1+R_2$, $L_{13}=R_1+R_6+R_7+R_3$, $L_{1\bar{1}}=R_1+R_6+R_5$,
etc. Then we can rewrite $L_f$ in the 
expectation values of the string potential
$\langle\Phi|H_f|\Phi\rangle$ 
with respect to a meson system($\Phi_{q\bar{q}}$), 
a three-quark system($\Phi_{q^3}$), and the pentaquark states
$\Phi_{(e)}$, $\Phi_{(f)}$, $\Phi_{(d)}$, as follows:
\begin{eqnarray}
L_f&=& L_{12} 
\quad {\rm in \ }\langle\Phi_{q\bar{q}}|H_f|\Phi_{q\bar{q}}\rangle,\\
L_f&=& \frac{1}{2}(L_{12}+L_{23}+L_{31})
\quad {\rm in \ }\langle\Phi_{q^3}|H_f|\Phi_{q^3}\rangle,\\
L_f&=& \frac{1}{2}(L_{12}+L_{34})+
\frac{1}{8}(L_{13}+L_{14}+L_{23}+L_{24})
+\frac{1}{4}(L_{\bar{1}1}+L_{\bar{1}2}+L_{\bar{1}3}+L_{\bar{1}4})
\quad {\rm in \ }\langle\Phi_{(e,f)}|H_f|\Phi_{(e,f)} \rangle,\\
L_f&=& \frac{1}{2}(L_{12}+L_{23}+L_{31})+L_{\bar{1}4}
\quad {\rm in \ }\langle\Phi_{(d)}|H_f|\Phi_{(d)}\rangle.
\end{eqnarray}
In the practical calculation, we approximate 
the minimum length of the flux tubes $L_f$ by
a linear combination of two-body distances $r_{ij}$ between the 
$i$-th (anti)quark the $j$-th (anti)quark as, 
\begin{eqnarray}
\label{eq:meson}
L_f&\approx& r_{12} 
\quad {\rm in \ }\langle\Phi_{q\bar{q}}|H_f|\Phi_{q\bar{q}}\rangle,\\
\label{eq:3q}
L_f&\approx& \frac{1}{2}(r_{12}+r_{23}+r_{31})
\quad {\rm in \ }\langle\Phi_{q^3}|H_f|\Phi_{q^3}\rangle,\\
\label{eq:5q}
L_f&\approx& \frac{1}{2}(r_{12}+r_{34})+
\frac{1}{8}(r_{13}+r_{14}+r_{23}+r_{24})
+\frac{1}{4}(r_{\bar{1}1}+r_{\bar{1}2}+r_{\bar{1}3}+r_{\bar{1}4})
\quad {\rm in \ }\langle\Phi_{(e,f)}|H_f|\Phi_{(e,f)} \rangle,\\
\label{eq:mb}
L_f&\approx& \frac{1}{2}(r_{12}+r_{23}+r_{31})+r_{\bar{1}4}
\quad {\rm in \ }\langle\Phi_{(d)}|H_f|\Phi_{(d)}\rangle.
\end{eqnarray}
It is clear that the above equations are obtained by approximating
the path length $L_{ij}$ with the distance $r_{ij}$ as 
$L_{ij}\approx r_{ij}$ for all $qq$ and $q\bar{q}$ pairs.
In the meson system, it is clear that 
Eq.\ref{eq:meson} gives the exact $L_f$ value.
The approximation, Eq.\ref{eq:3q}, for $3q$-baryons
is used in Ref.\cite{carlson} and has been proved to be a
good approximation.
We note that the confinement is reasonably realized by the 
approximation in Eq.\ref{eq:5q} for $\Phi_{(e,f)}$ as follows.
The flux-tube configuration (e)(or (f)) consists of seven bonds and 
three junctions. In the limit that the length($R_i$) of any $i$-th 
bond becomes much larger than other bonds, 
the string potential $\langle H_f \rangle$ 
approximated by Eq.\ref{eq:5q} behaves as a 
linear potential $\sigma R$.
It means that all the quarks and anti-quarks are 
bound by the linear potential with the tension $\sigma$.
In that sense, the approximation in Eq.\ref{eq:5q} for the connected
flux-tube configurations is regarded as a natural extension of the 
approximation(Eq.\ref{eq:3q}) for $3q$-baryons. 
It is convenient to introduce an operator
${\cal O}\equiv -\frac{3}{4}\sigma\sum_{i<j}F_i F_j r_{ij}-M^0$.  
One can easily prove that the above approximations, 
\ref{eq:meson},\ref{eq:3q},\ref{eq:5q},\ref{eq:mb}, are equivalent to 
$\langle\Phi|H_f|\Phi\rangle\approx \langle\Phi|{\cal O}|\Phi\rangle$
within each of the flux-tube configurations 
because the proper factors arise from $F_i F_j$ 
depending on the color configurations of the corresponding $qq$
(or $q\bar{q}$) pairs.

\begin{figure}
\noindent
\epsfxsize=0.8\textwidth
\centerline{\epsffile{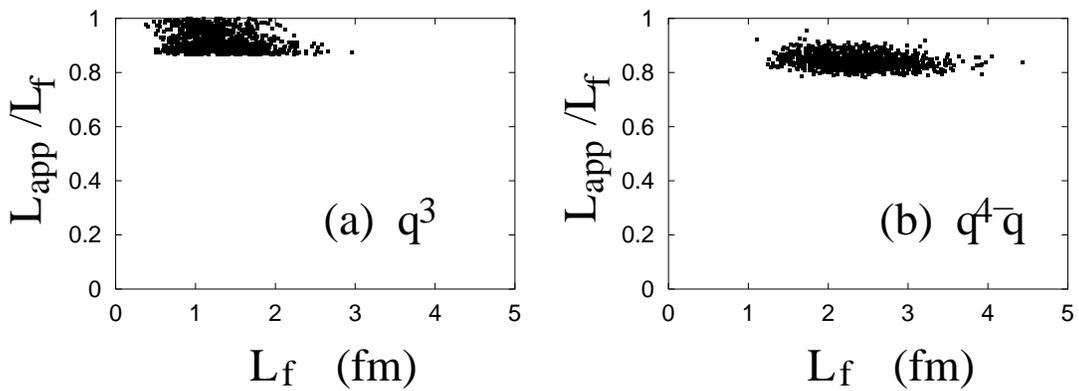}}
\ \\
\caption{\label{fig:flux-lf}
The ratio $L_{app}/L_f$ for the approximated tube length $L_{app}$
and the exact tube length $L_f$ in a $q^3$ system and a $[qq][qq]\bar{q}$ 
system. The quark positions ${\bf r}_i$ are randomly 
chosen in Gaussian deviates with the 
probability $\rho=\exp(-r_i^2/b^2)$, and $(L_f,L_{app}/L_f)$ values 
for 1000 samplings are plotted.}
\end{figure}

In order to see the accuracy of the approximations Eqs.\ref{eq:3q} 
and \ref{eq:5q}, 
we calculate the ratio of the approximated length $L_{app}$
to the exact $L_f$ in a simple quark distribution with 
Gaussian form which imitates the model wave function of the present
calculation.
Figure \ref{fig:flux-lf} shows the ratio $L_{app}/L_f$ 
in a $q^3$ system and a $[qq][qq]\bar{q}$ system. 
The quark positions ${\bf r}_i$ are randomly chosen in Gaussian 
deviates with the probability $\rho=\exp(-r_i^2/b^2)$, 
and $(L_f,L_{app}/L_f)$ values for 1000 samplings are plotted.
We use the same size parameter $b$ as that of the single-particle Gaussian
wave function in the present model explained later.
Comparing Figs.\ref{fig:flux-lf}(a) 
with \ref{fig:flux-lf}(b), we found that
the $L_{app}/L_f$ ratio for the $[qq][qq]\bar{q}$ system 
is about 10\% smaller than that for the $q^3$ system.
Since the zero-point energy $M_0$ in the string potential 
is adjusted in each of the $q^3$ and the $[qq][qq]\bar{q}$,
this underestimation should relate only to the relative energy
of the string potential in each system, and may give a minor effect 
on the level structure of the pentaquark.

We solve the eigenstates of the Hamiltonian with a variational method
in the AMD model space\cite{ENYObc,AMDrev}. 
We take a base AMD wave function in a quark model 
as follows. 
\begin{eqnarray}\label{eq:AMD}
&\Phi({\bf Z})=(1\pm P) {\cal A}
\left[\phi_{{\bf Z}_1}\phi_{{\bf Z}_2}\cdots
\phi_{{\bf Z}_{N_q}} X \right],\\
&\phi_{{\bf Z}_i}=\left(\frac{1}{\pi b^2}\right)^{3/4}
\exp\left[-\frac{1}{2b^2}({\bf r}-\sqrt{2}b
{\bf Z}_i)^2+\frac{1}{2}{\bf Z}^2_i\right],
\end{eqnarray}
where $1\pm P$ is the parity projection operator, ${\cal A}$ is the
anti-symmetrization operator,  
and the spatial part $\phi_{{\bf Z}_i}$ of 
the $i$-th single-particle wave function
given by a Gaussian whose center is located at ${\bf Z}_i$ in the phase space. 
$X$ is the spin-isospin-color function.
For example, in case of the proton, 
$X$ is given as
$X=\left(|\uparrow\downarrow\uparrow\rangle_S 
-|\uparrow\uparrow\downarrow\rangle_S \right) \otimes 
 |uud\rangle \otimes \epsilon_{abc}|abc\rangle_C $.
Here, $|m\rangle_S (m=\uparrow,\downarrow)$ 
is the intrinsic-spin function 
and $|a\rangle_C (a=1,2,3)$ expresses the color function.
Thus, the wave function of the $N_q$ quark system 
is described by the complex variational parameters, 
${\bf Z}=\{{\bf Z}_1,{\bf Z}_2,\cdots, {\bf Z}_{N_q}\}$. 
By using the frictional cooling method \cite{ENYObc}
the energy variation is performed 
with respect to ${\bf Z}$.

For the pentaquark system ($uudd\bar{s})$,
\begin{equation}
X=\sum_{m_{1},m_{2},m_{3},m_{4},m_{5}} 
c_{m_{1}m_{2}m_{3}m_{4}m_{5}}
|m_{1}m_{2}m_{3}m_{4}m_{5}\rangle_S 
\otimes \{ |udud\bar{s}\rangle {\rm \ or\ }
 |uudd\bar{s}\rangle \}\otimes  
\epsilon_{abg}\epsilon_{ceh}\epsilon_{ghf}|abce\bar{f}\rangle_C,
\end{equation}
where $|udud\bar{s}\rangle$ and $|uudd\bar{s}\rangle$ 
correspond to the configurations $[ud][ud]\bar{s}$ and 
$[uu][dd]\bar{s}$ in Fig.\ref{fig:flux}, respectively.
Since we are interested in the confined states, we do not 
use the meson-baryon states, $(qqq)_{1}(q\bar{q})_{1}$.
This assumption of decoupling of the 
reducible and irreducible configurations of the flux tubes
can be regarded as a kind of bound-state approximation.
The decoupling of the different flux-tube configurations can be
characterized by the suppression factor $\epsilon$ from 
the transition of the gluon field in the non-diagonal matrix elements 
$\epsilon\langle \Phi_1|{\cal O}|\Phi_2 \rangle$.
In a simple flux-tube model, $\epsilon$ is roughly estimated by the
area $\Delta_s$ swept by the tubes when moving from one configuration into
the other configuration as $\epsilon \sim \exp(-\sigma \Delta_s)$.
We make an estimation of the expectation value of $\exp(-\sigma \Delta_s)$ 
by assuming a simple quark distribution with 
Gaussian form which imitates the model wave function 
in the same way as the evaluation of the $L_{app}/L_f$.
The suppression factor $\epsilon$ among the configurations
$[ud][ud]\bar{s}$, $[uu][dd]\bar{s}$, and $(qqq)_{1}(q\bar{q})_{1}$
is estimated to be $\epsilon^2 \alt 1/10$ 
within the present model space. Therefore, we consider that 
the present assumption of the complete decoupling $\epsilon=0$ 
in the energy variation is acceptable in first order calculations.

The coefficients $c_{m_{1}m_{2}m_{3}m_{4}m_{5}}$ for the
spin function are determined by diagonalization of Hamiltonian and norm
matrices.
After the energy variation with respect to the $\{{\bf Z}\}$ and 
$c_{m_{1}m_{2}m_{3}m_{4}m_{5}}$, the intrinsic-spin and parity $S^\pi$ 
eigen wave function $\Phi({\bf Z})$ for the lowest state 
is obtained for each $S^\pi$. 
In the AMD wave function,
the spatial wave function is given by multi-center Gaussians.
When the Gaussian centers are located in some groups,  
the wave function describe the multi-center cluster structure and is
equivalent to the Brink model wave function(a cluster model often used  
in nuclear structure study)
\cite{Brink,Horiuchi-rev}.
On the other hand, because of the antisymmetrization, it can also represent 
shell-model wave functions when all the Gaussian centers are located 
near the center of the system\cite{Brink,Horiuchi-rev}.
In nuclear structure study, it has been already proved that 
the AMD is one of powerful tools 
due to the flexibility of the wave function\cite{AMDrev}.
In general, the relative motions in the AMD 
are given by such Gaussian forms as
$\exp[-\nu'({\bf x}-{\bf R})^2]$ where ${\bf x}$ is a Jacobi coordinate
and ${\bf R}$ is given by a linear combination of the Gaussian centers
$\{{\bf Z}_1,{\bf Z}_2,\cdots, {\bf Z}_{N_q}\}$.
Here we explain the details of the relative motion in a simple case of a 
two-body cluster structure in a $N_q=5$ system.
If the Gaussian centers are located in two groups
as ${\bf Z}_1={\bf Z}_2={\bf Z}_3={\bf Q}_1/\sqrt{2}b$ and 
${\bf Z}_4={\bf Z}_5={\bf Q}_2/\sqrt{2}b$, and if each group does not contain 
identical particles, the wave function expresses the two-body cluster state,
where each cluster is the harmonic oscillator $0s$-orbital state, 
$(0s)^{2,3}$, with zero orbital-angular momentum.
The inter-cluster motion ${\cal X}$ is given as 
${\cal X}({\bf x},{\bf R},\nu')=\exp[-\nu'({\bf x}-{\bf R})^2]$,
where $\nu'=\frac{3}{5b^2}$, ${\bf R}={\bf Q}_2-{\bf Q}_1$ 
and ${\bf x}$ is the relative coordinate between the clusters. 
In the partial wave expansion of the inter-cluster motion ${\cal X}$, 
\begin{eqnarray}
&{\cal X}({\bf x},{\bf R},\nu')=\exp[-\nu'({\bf x}-{\bf R})^2],\nonumber\\
&=\sum_L 4\pi i_L(2\nu'Rx)e^{-\nu'(x^2+R^2)}\sum_M
Y_{LM}(\hat x)Y^*_{LM}(\hat R),
\end{eqnarray}
where $i^L$ is the modified spherical Bessel function,
it is found that the wave function contains higher orbital-angular momentum 
$L$ components in general. However, in case of $\nu' R^2\le O(1)$, 
the wave function is dominated by the lowest $L$ component
since the $L$ components rapidly decrease with the increase of $L$.
As a result, the even-parity $S^{\pi=+}$ $3^q$ and 
odd-parity $q^4\bar{q}$ states 
are almost the $L=0$ eigen states, while 
the odd-parity $3^q$ and even-parity $q^4\bar{q}$ states 
are nearly the $L=1$ eigen states.
(The $q^4\bar{q}$ contains an odd intrinsic parity of the $\bar{q}$
in addition to the parity of the spatial part.)
Therefore, we do not perform the 
explicit $L$-projection in present calculation for simplicity.
We have actually checked that the obtained wave functions are almost the 
$L$-eigen($L=0$ or $1$) states and higher $L$ components are minor
in most of the $q^3$ and $q^4\bar{q}$ states. 

In the present wave function 
we do not explicitly perform the isospin projection, however, 
the wave functions obtained by energy variation 
are found to be approximately isospin-eigen states
in most of the low-lying states of the $q^3$ and $q^4\bar{q}$
due to the color-spin symmetry.

In the numerical calculation, the linear and Coulomb potentials
are approximated by seven-range Gaussians.
We use the following parameters,
\begin{eqnarray}
& \alpha_c =1.05, \nonumber\\
& \Lambda =0.13\ {\rm fm}, \nonumber\\
& m_q =0.313\ {\rm GeV}, \nonumber\\
& m_s =0.513\ {\rm GeV}, \nonumber\\
& {\sigma} =0.853\ {\rm GeV/fm}. 
\end{eqnarray}
Here, the quark-gluon coupling constant $\alpha_c$ is 
chosen so as to fit the $N$ and $\Delta$ 
mass difference.
The string tension $\sigma$ is
adopted to adjust the excitation energy of $N^*(1520)$.
The size parameter $b$ is chosen to be $0.5$ fm.

\section{Results}\label{sec:results}
In table.\ref{tab:3q}, we display the calculated energy of 
$q^3$ states with $S^\pi=1/2^+$($N$), $S^\pi=3/2^+$($\Delta$), 
$S^\pi=1/2^-$($N^*$).
The zero-point energy $M^0$ of the string 
potential is chosen to be $M^0_{q^3}=972$ MeV to fit the masses of 
$q^3$ systems, $N$, $N^*$ and $\Delta$. The calculated masses for $\Lambda$
with $S^\pi=1/2^-$ and $1/2^+$ correspond to the experimental data of 
$\Lambda(1115)$ and $\Lambda^*(1670)$.
The contributions of the kinetic and each 
potential terms are consistent with the results of  the
Ref.\cite{carlsonb}.
We checked that the obtained states are almost eigen states of the
angular momentum $L$ and the $L$ projection 
gives only minor effects on the energy.

\begin{table}
\caption{Calculated masses (GeV) of the $q^3$ systems.
The expectation values of the 
kinetic, string, Coulomb and color-magnetic
terms are also listed.}
\begin{center}
\begin{tabular}{rrrrrr}
 & $(uud)_{1}$ & $(uud)_{1}$ & $(uuu)_{1}$ &$(uds)_1$ & $(uds)_1$ \\
$S^\pi$ & $\frac{1}{2}^+$ & $\frac{1}{2}^-$ & $\frac{3}{2}^+$
& $\frac{1}{2}^+$ & $\frac{1}{2}^-$ \\
\hline
Kinetic($H_0$) &	1.74 	&	1.87 	&	1.66 	& 1.93 &  2.09\\
String($H_F$) &	0.02 	&	0.27 	&	0.07  & 0.03 & 0.25	\\
Coulomb&	$-$0.65 	&	$-$0.52 
	&	$-$0.62  & $-$0.65 & $-$0.53	\\
Color mag.&	$-$0.17 	&	$-$0.09 	&	0.14 & $-$0.16 & $-$0.14	\\
\hline
$E$ &	0.94 	&	1.52 	&	1.24 	& 1.14 & 1.67 \\
\hline
exp. (MeV)& $N(939)$ & $N^*(1520),N^*(1535)$ & $\Delta(1232)$ & $\Lambda(1115)$
& $\Lambda(1670)$ \\  
\end{tabular}
\end{center}\label{tab:3q} 
\end{table}

Now, we apply the AMD method to the $uudd\bar{s}$ system.
For each spin parity, we calculate energies of the  
 $[ud][ud]\bar{s}$ and 
$[uu][dd]\bar{s}$ states
and adopt the lower one.
In table.\ref{tab:5q}, the calculated results are shown.
We adjust the zero-point energy of the string potential $M_0$
as $M^0_{q^4\bar{q}}=2385$ MeV 
to fit the absolute mass of the recently observed $\Theta^+$.
This $M^0_{q^4\bar{q}}$ for pentaquark system is chosen
independently of $M^0_{q^3}$ for $3q$-baryon.
If $M^0_{q^4\bar{q}}=\frac{5}{3}M^0_{q^3}$ is assumed 
as Ref.\cite{carlsonb}, the calculated mass of the pentaquark
is around 2.2 GeV, which is consistent with the result of 
Ref.\cite{carlsonb}.

The most striking point in the results is that 
the $S^\pi=3/2^-$ and $S^\pi=1/2^+$ states nearly degenerate
with the $S^\pi=1/2^-$ states.
The $S^\pi=1/2^+$ correspond to $J^\pi=\{1/2^+,3/2^+\}$ ($S=1/2,L=1$), 
and the $S^\pi=3/2^-$ is 
$J^\pi=3/2^-$($S=3/2,L=0$).
The lowest $S^\pi=1/2^-$($J^\pi=1/2^-,L=0$) state appears just below the
$S^\pi=3/2^-$ and the second $S^\pi=1/2^-$($J^\pi=1/2^-,L=0$) state is 
at the same energy as the $S^\pi=1/2^+$($J^\pi=1/2^+,3/2^+$,$L=1$) states. 
However these $J^\pi=1/2^-$ states, as we discuss later, are expected to be 
much broader than other states. 
The $J^\pi=1/2^+$ and $3/2^+$ exactly degenerate
in the present Hamiltonian which does not contain the spin-orbit force. 
Other spin-parity states are much higher than these
low-lying states.

\begin{table}
\caption{Calculated masses(GeV) of the $uudd\bar{s}$ system.
$M^0_{q^4\bar{q}}$=2385 MeV is used to adjust 
the energy of the lowest state to the observed mass.
The expectation values of the 
kinetic, string, Coulomb, color-magnetic terms, and that 
of the color-magnetic term in $q\bar{q}$ pairs are listed.
In addition to the lowest $1/2^-$ state with the 
$[uu][dd]\bar{s}$ configuration, we also show the results of
the $1/2^-$ state with $[ud][ud]\bar{s}$ configuration, which lies
in the low-energy region.}
\begin{tabular}{rrrrrrrr}
& $[uu][dd]\bar{s}$
& $[ud][ud]\bar{s}$
& $[ud][ud]\bar{s}$
& $[ud][ud]\bar{s}$
& $[uu][dd]\bar{s}$
& $[ud][ud]\bar{s}$
& $[ud][ud]\bar{s}$
\\
$S^\pi$ & $\frac{1}{2}^-$& $\frac{3}{2}^-$& $\frac{1}{2}^+$& $\frac{1}{2}^-$&
 $\frac{5}{2}^-$
& $\frac{3}{2}^+$& $\frac{5}{2}^+$\\
\hline
Kinetic($H_0$) &3.23 &3.22 	&3.36 	&3.19 	&3.19 	&3.36 	&3.33 	\\
String($H_F$) & $-0.67$ &$-$0.66 	&$-$0.55&$-$0.64&$-$0.64&$-$0.56&$-$0.54 \\
Coulomb&$-$1.05&$-$1.04&$-$0.99&$-$1.03&$-$1.03&$-$0.99&$-$0.98\\
Color mag.&$-0.01$ &0.01 	&$-$0.25&0.04 	&0.19 	&$-$0.06 &0.17 	\\
$q\bar{q}$Color mag. &$-0.06$ &$-$0.01&0.00 	&0.02 	&0.06 	&0.02 	&0.04 	\\
\hline
$E$ & 1.50 &1.53 	&1.56 	&1.56 	&1.71 	&1.75 	&1.98 	\\
\end{tabular}\label{tab:5q}
\end{table}

\section{discussion}\label{sec:discuss}

In this section, we analyze the structure of the obtained 
low-lying states of the $uudd\bar{s}$ system, 
and discuss the level structure
and the width for $KN$ decays.

\subsection{Structure of low-lying states}
We analyze the spin structure of these states, and found that
the $J^\pi=\{1/2^+,3/2^+\}$ states consist of 
two spin-zero $[ud]$-pairs, while 
the $J^\pi=3/2^-$ contains
of a spin-zero $[ud]$-pair and a spin-one
$[ud]$-pair. 
Here we call the color anti-triplet $qq$ pair with 
the same spatial single-particle wave functions
as a $[qq]$-pair and note a spin $S$ $[qq]$-pair as 
$[qq]^S$. 
Since the $[ud]^{0}$-pair has the isospin $I=0$ and 
the $[ud]^{1}$-pair has the isospin $I=1$ 
because of the color asymmetry, the 
$J^\pi=3/2^-$ state is isovector while the 
lowest even-parity $J^\pi=\{1/2^+,3/2^+\}$ states are isoscalar.
The $J^\pi=1/2^+$ state corresponds to the $\Theta^+$(1530) 
in the flavor ${\overline{10}}$-plet 
predicted by Diakonov et al.\cite{diakonov}.
It is surprising that the odd-parity state, $J^\pi=3/2^-$ 
has the isospin $I=1$, which means that this state is a 
member of the flavor ${27}$-plet and belong to a 
new family of $\Theta$ baryon. 
We denote the $J^\pi=\{1/2^+,3/2^+\},
I=0$ states by $\Theta^+_0$, and the $J^\pi=3/2^-,I=1$ state 
by $\Theta^+_1$.
The mass difference $E(\Theta^+_0)-E(\Theta^+_1)$ is about 30 MeV.
In the energy region compatible to 
the $J^\pi=\{1/2^+,3/2^+\}$ and $J^\pi=3/2^-$ states, 
there appear two $J^\pi=\{1/2^-\}$ states. The lowest one
is the $[uu][dd]\bar{s}$ state with $[uu]^{1}$ and $[dd]^{1}$ pairs,
while the higher one is the $[ud][ud]\bar{s}$ with 
$[ud]^{0}$ and $[ud]^{1}$ pairs. The former is the isospin symmetric 
state and is dominated by $I=0$ component. The latter is isovector and
is regarded as the spin $S$-partner of the $J^\pi=3/2^-$ state.
The $J^\pi=1/2^-$ state is the lowest in the $uudd\bar{s}$ system.
We, however, consider this state not to be the observed $\Theta^+$
because its width should be broad as discussed later.

Although it is naively expected that unnatural spin parity states 
are much higher than the natural spin-parity $1/2^-$ state, 
the present results show the abnormal 
level structure of the $(udud\bar{s})$ system, where 
the high spin $J^\pi=3/2^-$ state and the 
unnatural parity $J^\pi=\{1/2^+,3/2^+\}$ states nearly degenerate
just above the $J^\pi=1/2^-$ state. 
By analysing the details of these states,
the abnormal level structure can be easily understood with a simple picture
as follows.
As shown in table.\ref{tab:5q}, 
the $J^\pi=\{1/2^+,3/2^+\}(L=1)$ states have larger
kinetic and string energies
than the $J^\pi=3/2^-(L=0)$ and $J^\pi=1/2^-(L=0)$
states, while the former states
gain the color-magnetic interaction. It indicates that the degeneracy of 
parity-odd states and parity-even states 
is realized by the balance of the loss of the kinetic and 
string energies and the gain of the color-magnetic interaction.
In the $J^\pi=\{1/2^+,3/2^+\}$ and the $J^\pi=1/2^-,3/2^-$ states, 
the competition of the energy loss and gain 
can be understood by Pauli principle 
from the point of view of the $[qq]$-pair structure as follows. 
As already mentioned by Jaffe and Wilczek\cite{jaffe}, the relative motion 
between two $[qq]^{0}$-pairs must have the odd parity ($L=1$)
because the $L=0$ is forbidden between the two identical $[qq]$-pairs
due to the color antisymmetry. In the $J^\pi=3/2^-$ state and the second
$J^\pi=1/2^-$ state, one of $[ud]^{0}$-pairs is broken to be a 
$[ud]^{1}$-pair and the $L=0$ is allowed because two diquarks 
are not identical. 
The $L=0$ is energetically favored in the kinetic and string terms, and 
the energy gain cancels the color-magnetic energy loss of a 
$[ud]^{1}$-pair. 
Also in the lowest $J^\pi=1/2^-$ state,
the competition of energy loss and gain is similar
as each contribution of the kinetic, 
string and potential energies in the lowest $J^\pi=1/2^-$ state 
is almost the same as those in the $J^\pi=3/2^-$ and the second
$J^\pi=1/2^-$(table \ref{tab:5q}). 
It means that the gain of the kinetic energy of the $L=0$ state compete
with the color-magnetic energy loss in the lowest $J^\pi=1/2^-$ 
as well as the $J^\pi=3/2^-$ and the second $J^\pi=1/2^-$.

\begin{figure}
\noindent
\epsfxsize=0.4\textwidth
\centerline{\epsffile{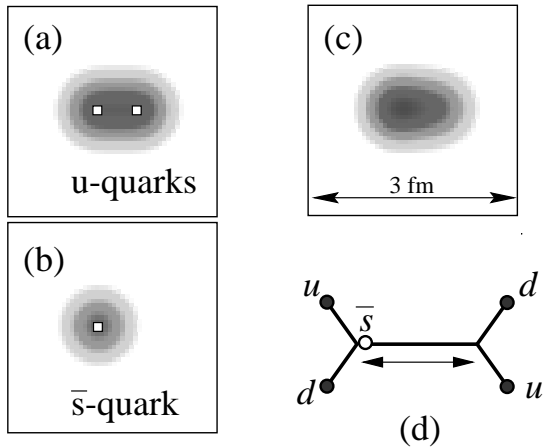}}
\  \\
\caption{\label{fig:5q}
The $q$ and $\bar{q}$ density distribution in the 
$J^\pi=1/2^+,3/2^+(S=1/2,L=1)$ states of the $uudd\bar{s}$ system.
The $u$ density (a), $\bar{s}$ density (b), and total quark-antiquark density
(c) of the intrinsic state before parity projection are shown. 
The schematic figure of the corresponding flux-tube configuration
is illustrated in (d). Open squares in (a) and (b), indicate 
the positions of Gaussian centers ${\rm Re}[\protect\sqrt{2}b{\bf Z}_{i}]$ 
for
the $i$-th single-particle wave functions.
}
\end{figure}

We should stress that the existence of two spin-zero $ud$-diquarks in
the $J^\pi=\{1/2^+,3/2^+\}$ states predicted 
by Jaffe and Wilczek\cite{jaffe} is actually confirmed 
in the present calculations without {\it a priori} assumptions
for the spin and spatial configurations.
In fact, the component 
with two spin-zero $[ud]$-pairs is 97\% in the present 
$J^\pi=\{1/2^+,3/2^+\}$ state.
In Fig.\ref{fig:5q}, we show the quark and 
anti-quark density distribution in the
$J^\pi=\{1/2^+,3/2^+\}$ states and display the centers of Gaussians 
for the single-particle wave functions.
In the intrinsic wave function, Gaussian centers for two 
$[ud]^{0}$-pairs are located far from each other with the 
distance about 0.6 fm. It indicates 
the spatially developed diquark-cluster
structure, which means the spatial and spin correlations in 
each $[ud]$-pair.
It is found that the center of the $\bar{s}$ stays at the same point of 
that of one $[ud]^{0}$, as 
${\bf Z}_1={\bf Z}_2=\frac{3}{5\sqrt{2}b}{\bf Q}_{12}$ and 
${\bf Z}_3={\bf Z}_4={\bf Z}_5=-\frac{2}{5\sqrt{2}b}{\bf Q}_{12}$ 
where $\{{\bf Z}_1,{\bf Z}_2,\cdots,{\bf Z}_5\}$ are the Gaussian centers 
in Eq.\ref{eq:AMD} and $|{\bf Q}_{12}|\sim 0.6$ fm.
As a result, we found the spatial development of $ud$-$uds$ clustering 
and a parity-asymmetric shape in the intrinsic state 
before parity projection(Fig.\ref{fig:5q}). 
As explained in \ref{sec:framework}, the wave function is 
equivalent to the $[ud]^{0}$-$[ud]^{0}\bar{s}$ cluster 
wave function in Brink model\cite{Brink} with $L=1$ relative motion. 
After the parity projection, the $\bar{s}$ is exchanged between 
two diquarks. In contrast to the spatially developed cluster structure in
the even-parity state, the odd-parity states $J^\pi=1/2^-, 3/2^-$  
are almost the spatially symmetric $(0s)^5$ states with spherical shapes. 

As mentioned before, the degeneracy of the even-parity states and 
the odd-parity states originates in the balance of the $L=1$ excitation 
energy and the energy gain of the color-magnetic interaction.
Here we consider the $L=1$ excitation energy $\Delta E(L=1)$ 
as the total energy lose in the
kinetic, string and Coulomb terms.
It is important that $\Delta E(L=1)\sim 0.3$ GeV in the pentaquark 
is much smaller than $\Delta E(L=1)\sim 0.5$ GeV in the nucleon 
system. The reason for the relatively small $\Delta E(L=1)$ in the pentaquark
can be easily understood by the $ud$-$uds$ cluster structure.
In the two-body cluster state with the $L=1$ relative motion,
the $\Delta E(L=1)$ is roughly estimated by the reduced mass 
$\mu=A_1A_2/(A_1+A_2)$ 
of two clusters, as is given 
as $\Delta E(L=1)\propto \frac{1}{\mu}$ ($A_1$ and $A_2$
are the masses of the clusters).
In the nucleon, $\mu=\frac{2}{3} m_q$ is obtained from 
the $ud$-$u$ cluster structure in the $J^\pi=1/2^-(L=1)$ state, 
while $\mu \sim \frac{6}{5} m_q$ for the pentaquark system 
is found in the $ud$-$uds$ clustering. The reduced mass in the pentaquark 
is $9/5$ times larger than that in the nucleon system, therefore,
$\Delta E(L=1)$ should be smaller in the pentaquark than in the nucleon 
by the factor 
$5/9$. This factor is consistent with the present $\Delta E(L=1)$ values.

We give a comment on the $LS$-splitting between $J^\pi=1/2^+$ and 
$3/2^+(S=1/2,L=1)$.
In the present calculation, where the spin-obit force is omitted, 
the $J^\pi=1/2^+$ and $3/2^+$ states exactly degenerate. Even if we introduce
the spin-orbit force into the Hamiltonian, 
the $LS$-splitting should not be large
in this diquark structure because the effect of the spin-orbit force from the 
spin-zero diquarks is very weak as discussed in Ref.\cite{Close}. 

We remark that the $[ud]^{0}$-$[ud]^{0}\bar{s}$ cluster
structure in the present result is different from the 
diquark-triquark structure proposed by
Karliner and Lipkin \cite{karliner} because the $ud\bar{s}$-triquark
in Ref.\cite{karliner} 
is the $(us)^{S=1}_6\bar{s}$ with the color-symmetric spin-one 
$ud$-diquark. In the $(us)^{S=1}_6\bar{s}$-triquark, the $\bar{s}$ quark should
be tightly bound in the triquark due to the strong color-magnetic
interaction between $(us)^{S=1}_6$ and $\bar{s}$.
On the other hand, in the present $[ud]^{0}\bar{s}$-cluster, 
the $\bar{s}$ feels no strong color-magnetic interaction 
and is bound more weakly than in the $(us)^{S=1}_6\bar{s}$-triquark.
The color-6 flux tubes are not taken into account in the present framework
since they are excited. However, 
the $(us)^{S=1}_6\bar{s}$-triquark might
be possible if the short-range correlation in the triquark make the
flux-tube short enough to be excited into the color-6 
flux-tube.

\subsection{Width for $KN$ decays}

In the $\Theta^+$ $\rightarrow$ $KN$ decays,
it is important that the allowed decay mode 
in the $\Theta^+_1$($J^\pi=3/2^-$) is $D$ wave, 
which should make the $\Theta^+_1$ state 
narrower than the $\Theta^+_0$($J^\pi=1/2^+,3/2^+$) 
because of higher centrifugal barrier.
We estimate the $KN$-decay widths of these states by using a method 
of reduced width amplitudes\cite{Horiuchi-rev,Horiuchi73}. 
This method has been applied 
for the study of $\alpha$-decay width in the nuclear physics 
within bound state approximations. 
In this method, the decay width $\Gamma$
is estimated by the penetrability $P_L(k,a)$ of the barrier 
and the reduced width $\gamma^2(a)$ as a function of the 
threshold energy $E_{th}$ and the channel radius $a$,
\begin{eqnarray}
\Gamma=2P_L(k,a)\gamma^2(a),\nonumber\\
P_L(k,a)=\frac{ka}{j^2_L(ka)+n^2_L(ka)},\nonumber\\
\gamma^2(a)=\frac{\hbar^2}{2\mu a}S_{fac}(a),
\end{eqnarray}
where $\mu$ is the reduced mass, $k$ is the wave number 
$k=\sqrt{2\mu E_{th}/\hbar^2}$, and $j_L$($n_L$) is
the regular(irregular) spherical Bessel function.
$S_{fac}(a)$ is the probability of decaying particle 
at the channel radius $a$.
We define 
$\Gamma^0_L(a,E_{th})\equiv\frac{\hbar^2k}{\mu}\frac{1}{j^2_L(ka)+n^2_L(ka)}$,
then, the decay width can be rewritten in a simple form as  
$\Gamma=\Gamma^0_L\times S_{fac}$.
In the following discussion,
we choose the channel radius $a=1$ fm and $E_{th}=100$ MeV.
Since the transitions between the different flux-tube configurations, 
a confined state
$[ud][ud]\bar{s}$ and a decaying state
$(udd)_{1}(u\bar{s})_{1}$, are of higher order,
the $S_{fac}$ should be small in general when the suppression by the 
flux-tube transition is taken into account.
Here, we evaluate the maximum values of the 
widths for the $J^\pi=1/2^+, 3/2^+$ states with the method of the 
reduced width amplitudes, by using meson-baryon probability considering
only the simple overlap for the quark wave functions.

In case of even-parity $J^\pi=1/2^+, 3/2^+$ states, the $KN$ decay modes 
are the $P$-wave, which gives $\Gamma^0_{L=1}\approx 100$ MeV\ fm.
By assuming $(0s)^2$ and $(0s)^3$ harmonic-oscillator wave functions for 
$K^0$ and $p$, we calculate the overlap between the obtained pentaquark
wave function and the $K^0p$ state. As explained in the previous subsection,
the $J^\pi=1/2^+, 3/2^+$ states have 
the $ud$-$ud\bar{s}$ cluster structure where five Gaussian centers 
are written as ${\bf Z}_1={\bf Z}_2=\frac{3}{5\sqrt{2}b}{\bf Q}_{12}$ and 
${\bf Z}_3={\bf Z}_4={\bf Z}_5=-\frac{2}{5\sqrt{2}b}{\bf Q}_{12}$. 
We assume a simple $K^0p$ wave function as follows, 
\begin{eqnarray}
&\Phi_{K^0p}=(1+P) {\cal A}
\left[\phi_{{\bf Z}_1}\phi_{{\bf Z}_2}\cdots
\phi_{{\bf Z}_{N_q}} X \right],\\
&\phi_{{\bf Z}_i}=\left(\frac{1}{\pi b^2}\right)^{3/4}
\exp\left[-\frac{1}{2b^2}({\bf r}-\sqrt{2}b
{\bf Z}_i)^2+\frac{1}{2}{\bf Z}^2_i\right],
\end{eqnarray}
where the ${\bf Z}$a are chosen as 
${\bf Z}_1={\bf Z}_2={\bf Z}_3=a \frac{2}{5\sqrt{2}b}
{\bf Q}_{12}/|{\bf Q}_{12}|$,
 ${\bf Z}_4={\bf Z}_5=- a \frac{3}{5\sqrt{2}b}
{\bf Q}_{12}/|{\bf Q}_{12}|$, 
and the spin-isospin-color wave function is taken to be
\begin{equation}
X=\sum_{m_{1},m_{2},m_{3},m_{4},m_{5}} 
c_{m_{1}m_{2}m_{3}m_{4}m_{5}}
|m_{1}m_{2}m_{3}m_{4}m_{5}\rangle_S 
\otimes |udud\bar{s}\rangle \otimes  
\epsilon_{abc}\delta_{ef}|abce\bar{f}\rangle_C.
\end{equation}
The same size parameter $b$ as that of the
pentaquark is used.
The coefficients $c_{m_{1}m_{2}m_{3}m_{4}m_{5}}$ for the
spin function are taken to express the $J^\pi=1/2^+$ proton and 
the pseudoscalar $K^0$ meson. 
The probability $S_{fac}$ is evaluated by the overlap with the obtained
$J^\pi=1/2^+,3/2^+$ wave function,
$S_{fac}=|\langle \Phi_{K^0p}|\Phi({\bf Z})\rangle|^2$. 
(The $\Phi_{K^0p}$ and $\Phi({\bf Z})$ are normalized.)
The probability $S_{fac}=0.034$
fm$^{-1}$ is evaluated by the overlap.
Roughly speaking, the main factors in this meson-baryon probability 
are the factor $\frac{1}{3}$ from the color configuration, 
the factor $\frac{1}{4}$ from the intrinsic spin part, and
the other factor which arises from the spatial overlap.
By using the probability $S_{fac}=0.034$, the $K^0p$ partial decay width 
is evaluated as $\Gamma < 3.4$ MeV. The 
$K^+n$ decay width is the same as that of the 
$K^0p$ decay, and the total width of the $J^\pi=1/2^+,3/2^+$ states 
is estimated to be $\Gamma < 7$ MeV. This is consistent with the
discussion in Ref.\cite{Carlson-width}. 

It is interesting that the $KN$ decay width of 
the $\Theta^+_1$($J^\pi=3/2^-$) is strongly suppressed by the 
$D$-wave centrifugal barrier, because lower spin ($S$-wave
and $P$-wave) decays are forbidden due to the conservation of spin and 
parity. Consequently, 
$\Gamma^0_{L=2}$ is $\approx 30$ MeV\ fm, which is much smaller 
than that for the $P$-wave case.
Moreover, the $\Theta^+_1(J^\pi=3/2^-)$ 
is the state with $S^\pi=3/2^-$ and $L=0$,
which has no overlap with the $KN$($S^\pi=1/2^-$ and $L=2$) states 
in the present calculation because the spin-orbit or tensor forces are
omitted.
If we introduce the spin-orbit or tensor forces, 
the $D$-state($S^\pi=1/2^-$ and $L=2$) will be slightly mixed 
into the $\Theta^+_1(J^\pi=3/2^-)$. 
However, the mixing component should be small because of the dominant 
central force in the potential.
In other words,  the $KN$ probability($S_{fac}$) 
in the $\Theta^+_1(J^\pi=3/2^-)$ state
is expected to be rather suppressed than that in the
$\Theta^+_0(J^\pi=1/2^+,3/2^+)$ states.  
Considering the suppression effects in both terms $\Gamma^0$ and
$S_{fac}$, the $J^\pi=3/2^-$ state should be extremely narrow.
If we assume the $S_{fac}$ in the $J^\pi=3/2^-$ to be half of that in
the $J^\pi=1/2^+, 3/2^+$ states,
the $KN$ decay width is estimated to be $\Gamma<1$ MeV.

Contrary to the narrow width of the $J^\pi=3/2^-$ state, 
the $J^\pi=1/2^-$ state should be much
broader than other states because $S$-wave($L=0$) decay is allowed 
and therefore the centrifugal barrier is absent.
We cannot evaluate the width of the $J^\pi=1/2^-$ states with the present
method, since the method of the reduced width amplitudes works 
only when there exist
barriers in the decaying channels.
If we adopt the theoretical width $\Gamma=1.1$ GeV for the 
$J^\pi=1/2^-$ states in \cite{Carlson-width} and the 
suppression factor $\epsilon^2\alt 1/10$ due to the string transition,
the width is evaluated to be $\Gamma\sim 100$ MeV, which is still too large
to describe that of the observed $\Theta^+$. 
We consider that the $J^\pi=1/2^-$ states may melt away due to the 
coupling with $KN$ continuum states with no centrifugal barrier.

Also in other spin-parity states, 
the coupling with the $KN$ continuum states is important
for more quantitative discussions on the widths.
We should point out that, in introducing the meson-baryon coupling, 
one should not treat only the quark degrees of freedom
but take into account the suppression due to the rearrangement of 
flux-tube topologies between the meson-baryon states and the confined states.

\section{Summary}\label{sec:summary}
We proposed a quark model in the framework of the AMD method,
and applied it to the $uudd\bar{s}$ system. 
The level structure of the $uudd\bar{s}$ system 
and the properties of the low-lying states were studied
within the model space of the $[qq][qq]\bar{q}$ configuration, where all the
(anti)quarks are connected by the color-$3$ flux tubes.
We predicted that the narrow 
$J^\pi=\{1/2^+$,$3/2^+$\}($\Theta_{0}$) 
and $J^\pi=3/2^-$ ($\Theta_{1}$) states
nearly degenerate with the $J^\pi=1/2^-$ states. 
The widths of the $J^\pi=\{1/2^+,3/2+\}$ states and the $3/2^-$
state are estimated 
to be $\Gamma < 7$ MeV and $\Gamma < 1$ MeV, respectively. On the other hand, 
the $J^\pi=1/2^-$ states should be broad, and we consider that they
may melt away due to the coupling with $KN$ continuum states 
with no centrifugal barrier. 
Two spin-zero diquarks are found in the $\{1/2^+,3/2+\}$ states, 
which confirms Jaffe-Wilczek picture.
We comment that the formation of two spin-zero diquarks does not always
occur in $J^\pi=\{1/2^+,3/2^+\}$ pentaquarks. For example, in case of 
the $ddss\bar{u}$ system, the diquark structure disappears. Instead, 
a $dss$-$d\bar{u}$ clustering appears in the $J^\pi=\{1/2^+,3/2^+\}$ 
$[ds][ds]\bar{u}$ states
because the color magnetic interaction is weaker for $ds$ pairs than for
$d\bar{u}$ pairs in OGE potential. 
In other words, the diquark structure is formed 
in such a certain pentaquark as the $\Theta^+_0$ 
due to the strong color-magnetic attraction between $ud$ quarks.
The degeneracy of the $J^\pi=1/2^-$, $3/2^-$, $1/2^+$ and $3/2^+$ states
is realized by the balance of 
the kinetic and string energies and the color-magnetic interaction. 
The origin of the novel level structure 
is due to the color structure in the confined five quark system 
bound by the connected flux-tubes.

The $J^\pi=\{1/2^+,3/2^+\}$$(\Theta^+_{I=0}$) states in the present 
results may be assigned to the experimentally observed $\Theta^+$, 
while $J^\pi=3/2^-$($\Theta_{I=1}$) is not observed yet.
One should pay attention to 
the properties of these states,
because the production rates depend 
on their spin, parity and widths. 
The existence of many narrow states,
$J^\pi=1/2^+$, $3/2^+$, and $3/2^-$, 
for the $\Theta^+_0$ and $\Theta^+_1$
may help to explain the inconsistent mass positions of 
the $\Theta^+$ among the different experiments.
Especially, the double peaks of the 
$J^\pi=1/2^+$ and $3/2^+$ states in the $\Theta^+_0$ are expected.
In the $\Theta^+_0$ peaks observed in the invariant mass 
or missing mass spectra, it is difficult to find the possible double peaks 
because the statistics and the resolutions are not 
enough\cite{leps,diana,clasa,clasb,saphir,itep,hermes,itep-2,zeus}.
The analysis of the $NK$ scattering\cite{Cahn} 
provided the upper limit $\Gamma <1$ MeV for the 
widths of each peaks. Considering the suppression factor due to the 
gluon transitions, the possibility of the double peaks
($J^\pi=1/2^+$ and $3/2^+$) suggested in the present works 
has not been excluded yet. 
We should comment that 
another explanation for the inconsistency of the experimental 
mass positions was suggested in Ref.\cite{Zhao}, where 
a systematic lowering in mass of $K^0p$ peaks relative to the
$K^+n$ was noted.
In the $I=1$ channel, there is no significant $\Theta^{++}$ signal 
in the experimental data of the invariant $K^+p$ mass 
in the photo-induced reactions\cite{saphir,zeus,clas-c}. 
It is important that 
the widths of the $J^\pi=3/2^-$ $\Theta_{I=1}$ should be about one order 
smaller than those of the $J^\pi=1/2^+$ and $3/2^+$ for the $\Theta^{+}_0$.
For the $\Theta^{++}$ search, it would be helpful to choose proper 
entrance and decay channels 
based on the further investigation of the production mechanism.
In order to compare the present findings with the experimental data 
is more details, further experimental data with high resolution 
and high statistics are required.

Finally, we would like to remind the readers that the 
absolute masses of the pentaquark in the present work are not predictions. 
We have an ambiguity of the zero-point energy of the string potential,
which depends on the flux-tube topology in each of meson, three-quark baryon, 
pentaquark systems. In the present calculation of the pentaquark, 
we phenomenologically adjust it to reproduce the observed mass of 
the $\Theta^+$. 
In order to predict absolute masses of unknown multiquarks 
with new flux-tube topologies, it is desirable to 
determine the zero-point energy more theoretically.

The authors would like to thank to T. Kunihiro, Y. Akaishi and H. En'yo 
for valuable discussions.
This work was supported by Japan Society for the Promotion of 
Science and Grants-in-Aid for Scientific Research of the Japan
Ministry of Education, Science Sports, Culture, and Technology.


\def\Ref#1{[\ref{#1}]}
\def\Refs#1#2{[\ref{#1}\ref{#2}]}
\def\npb#1#2#3{{Nucl. Phys.\,}{\bf B{#1}}\,(#3)\,#2}
\def\npa#1#2#3{{Nucl. Phys.\,}{\bf A{#1}}\,(#3)\,#2}
\def\np#1#2#3{{Nucl. Phys.\,}{\bf{#1}}\,(#3)\,#2}
\def\plb#1#2#3{{Phys. Lett.\,}{\bf B{#1}}\,(#3)\,#2}
\def\prl#1#2#3{{Phys. Rev. Lett.\,}{\bf{#1}}\,(#3)\,#2}
\def\prd#1#2#3{{Phys. Rev.\,}{\bf D{#1}}\,(#3)\,#2}
\def\prc#1#2#3{{Phys. Rev.\,}{\bf C{#1}}\,(#3)\,#2}
\def\prb#1#2#3{{Phys. Rev.\,}{\bf B{#1}}\,(#3)\,#2}
\def\pr#1#2#3{{Phys. Rev.\,}{\bf{#1}}\,(#3)\,#2}
\def\ap#1#2#3{{Ann. Phys.\,}{\bf{#1}}\,(#3)\,#2}
\def\prep#1#2#3{{Phys. Reports\,}{\bf{#1}}\,(#3)\,#2}
\def\rmp#1#2#3{{Rev. Mod. Phys.\,}{\bf{#1}}\,(#3)\,#2}
\def\cmp#1#2#3{{Comm. Math. Phys.\,}{\bf{#1}}\,(#3)\,#2}
\def\ptp#1#2#3{{Prog. Theor. Phys.\,}{\bf{#1}}\,(#3)\,#2}
\def\ib#1#2#3{{\it ibid.\,}{\bf{#1}}\,(#3)\,#2}
\def\zsc#1#2#3{{Z. Phys. \,}{\bf C{#1}}\,(#3)\,#2}
\def\zsa#1#2#3{{Z. Phys. \,}{\bf A{#1}}\,(#3)\,#2}
\def\intj#1#2#3{{Int. J. Mod. Phys.\,}{\bf A{#1}}\,(#3)\,#2}
\def\sjnp#1#2#3{{Sov. J. Nucl. Phys.\,}{\bf #1}\,(#3)\,#2}
\def\pan#1#2#3{{Phys. Atom. Nucl.\,}{\bf #1}\,(#3)\,#2}
\def\app#1#2#3{{Acta. Phys. Pol.\,}{\bf #1}\,(#3)\,#2}
\def\jmp#1#2#3{{J. Math. Phys.\,}{\bf {#1}}\,(#3)\,#2}
\def\cp#1#2#3{{Coll. Phen.\,}{\bf {#1}}\,(#3)\,#2}
\def\epjc#1#2#3{{Eur. Phys. J.\,}{\bf C{#1}}\,(#3)\,#2}
\def\mpla#1#2#3{{Mod. Phys. Lett.\,}{\bf A{#1}}\,(#3)\,#2}
\def\etal{{\it et al.}}

\end{document}